\documentclass[conference]{IEEEtran}
\usepackage{epsfig,psfrag}
\usepackage[latin9]{inputenc}
\usepackage{amsmath,amsfonts,amssymb,amsxtra,bm}
\usepackage{graphicx}
\usepackage{tikz}
\usepackage{color}
\usepackage{cite}
\usepackage{setspace}
\usepackage{tikz}
\usepackage{pgfplots}
\usepackage{caption}
\usepackage{notation}
\usepackage{cuted}
\usepackage{flushend}
\usepackage{lipsum}
\usepackage{bbm}

\newcommand{\ist}{\hspace*{.3mm}}
\newcommand{\rmv}{\hspace*{-.3mm}}

\newcommand{\be}{\begin{equation}}
\newcommand{\ee}{\end{equation}}
\newcommand{\iist}{\hspace*{1mm}}
\newcommand{\rrmv}{\hspace*{-1mm}}

\newcommand{\nn}{\nonumber}

\newcommand{\T}{\text{T}}

\newcommand{\va}[1]{#1}

\newcommand{\RFSst}{\Set{X}}
\newcommand{\RFSstR}{\RS{X}}	
\newcommand{\st}{\V{x}}
\newcommand{\stR}{\RV{x}}
\newcommand{\ex}{r}
\newcommand{\sd}{f}

\newcommand{\me}{z}
\newcommand{\meR}{\rv{z}}
\newcommand{\mev}{\V{z}}
\newcommand{\mevR}{\RV{z}}

\newcommand{\ass}{\va{a}}								
\newcommand{\assv}{\V{a}}								
\newcommand{\assR}{\rv{a}}							
\newcommand{\assvR}{\RV{a}}

\newcommand{\assw}{\beta}
\newcommand{\su}{p_{\text{S}}}
\newcommand{\de}{p_{\text{D}}}
\newcommand{\PHD}{\lambda}

\DeclareMathAlphabet{\mathpzc}{OT1}{pzc}{m}{it}

\allowdisplaybreaks
\pgfplotsset{compat=1.18}
\begin{document}



\title{Multiobject Tracking for \\[.3mm] Thresholded Cell Measurements \vspace{-3mm} }
\author{\IEEEauthorblockN{Thomas Kropfreiter\IEEEauthorrefmark{1}, Jason L. Williams\IEEEauthorrefmark{2}, and Florian Meyer\IEEEauthorrefmark{1}\vspace{2mm}}
\IEEEauthorblockA{\IEEEauthorrefmark{1}University of California San Diego, La Jolla, CA, USA (e-mail: tkropfreiter@ucsd.edu,\ist flmeyer@ucsd.edu)\vspace{1mm}}
\IEEEauthorblockA{\IEEEauthorrefmark{2}Whipbird Signals, Brisbane, Australia (jason@whipbird.au)\vspace{-8.5mm}}
\thanks{This work was supported by the Austrian Science Fund (FWF) under Grant J\,4726-N and the National Science Foundation (NSF) under CAREER award 2146261. }
}
\IEEEoverridecommandlockouts

\maketitle

\begin{abstract}
In many multiobject tracking applications, including radar and sonar tracking, after prefiltering the received signal, measurement data is typically structured in cells. The cells, e.g., represent different range and bearing values. However, conventional multiobject tracking methods use so-called point measurements. Point measurements are provided by a preprocessing stage that applies a threshold or detector and breaks up the cell's structure by converting cell indexes into, e.g., range and bearing measurements. We here propose a Bayesian multiobject tracking method that processes measurements that have been thresholded but are still cell-structured.
We first derive a likelihood function that systematically incorporates an adjustable detection threshold which makes it possible to control the number of cell measurements. We then propose a Poisson Multi-Bernoulli (PMB) filter based on the likelihood function for cell measurements.
Furthermore, we establish a link to the conventional point measurement model by deriving the likelihood function for point measurements with amplitude information (AM) and discuss the PMB filter that uses point measurements with AM. Our numerical results demonstrate the advantages of the proposed PMB filter for thresholded cell measurements compared to the conventional PMB filter for point measurements with and without AM. 

\vspace{1mm}
\end{abstract}

\begin{IEEEkeywords}
Multiobject tracking, 
multitarget tracking,
Poisson Multi-Bernoulli filtering,
random finite 
\vspace{-2.5mm}
sets.
\end{IEEEkeywords}

\vspace{2mm}
\section{Introduction}
\label{sec:Int}
Multiobject tracking (MOT) aims to estimate the time-dependent number and states of multiple objects based on data provided by one or more sensors. 
MOT algorithms use either sensor data in a preprocessed form or raw sensor data without any preprocessing.
In the first approach, known as \emph{detect-then-track} (DTT) MOT, a detector discards all measurements that are below a predefined threshold in order to reduce data flow and computational complexity \cite{Bar11,Cha11,Koc14,Bar95,For83,Rei79,Mah07,Mah03,Vo13,Wil:J15,Mey18Proc,Mey17MSBP,Kro16,KroMeyCorCarMenWil:C21,Kro19LMB,Kro22EffRFS,Gar18PMBM}. In the second approach, known as \emph{track-before-detect} (TBD) MOT, the tracker uses all the raw sensor data, which typically results in improved tracking performance at a higher computational complexity.

Irrespective of the measurement model that is in force, the multiobject state can either be modeled as a random vector \cite{Bar11,Cha11,Koc14,Bar95,For83,Rei79,Mey18Proc,Mey17MSBP,KroMeyCorCarMenWil:C21,Mey21EOT,LiLeiVenTuf:J22} or a random set, more precisely, a random finite set (RFS) \cite{Mah07,Mah03,Vo13,Wil:J15,Kro16,Kro19LMB,Kro22EffRFS,Gar18PMBM}.
A high-performing RFS-based MOT filter is the track-oriented marginal multi-Bernoulli/Poisson filter, which will be simply referred to as Poisson Multi-Bernoulli (PMB) filter in the following.
The PMB filter was both proposed for DTT MOT \cite{Wil:J15,Mey18Proc,Kro16,Gar18PMBM} and recently also for TBD MOT \cite{Kro21TBD}.
The PMB filter models the multiobject state as the union of a Poisson RFS and a multi-Bernoulli (MB) RFS.
Whereas the MB part of the PMB filter is used for the tracking of already detected objects, the Poisson part models objects that have not been detected so far. 
A modified version of the PMB filter proposed in \cite{Wil12} extends the use of the Poisson RFS from the modeling of undetected objects to the tracking of objects that are unlikely to exist by using the concept of recycling.
In the recycling step, Bernoulli components with a low existence probability are transferred from the MB part of the multiobject state to the Poisson part of the multiobject state instead of pruning them.   
The PMB filter can be implemented very efficiently using the framework of belief propagation (BP) \cite{Mey18Proc,Wil14} resulting in an overall linear complexity scaling.
The PMB filter has also been derived and extended to multiple sensors using the framework of factor graphs and BP \cite{Mey18Proc,Mey17MSBP,Mey21EOT,LiLeiVenTuf:J22}. It has been demonstrated that it can outperform existing DTT methods and that it is highly scalable in all relevant system parameters \cite{Wil12,Wil14,Wil:J15,Mey18Proc,Mey17MSBP,Kro16,KroMeyCorCarMenWil:C21,Mey21EOT,Kro21TBD}.

In this paper, we introduce a measurement model for thresholded cell measurements and propose a PMB filter based on this new model. 
The novel method can be interpreted as a tradeoff between the already existing DTT PMB and TBD PMB filters. 
In particular, as in DTT filtering, a threshold controls the number of considered measurements, but, as in TBD filtering, measurements still possess the cell structure of the original, unthresholded, measurements. 
In addition, this cell structure makes it possible to avoid the Poisson clutter approximation used by conventional DTT approaches. We also establish a link to the conventional likelihood function for point measurements with amplitude information (AM) and discuss the approximations applied in conventional point measurement models. Our numerical results indicate an improved tracking performance compared to the DTT PMB filter with and without AM in both tracking accuracy and runtime.

We will use the following basic notation. Random variables are displayed in sans serif, upright fonts; their realizations in serif, italic fonts. Vectors and matrices are denoted by bold lowercase and uppercase letters, respectively. For example, $\rv{x}$ is a random variable and $x$ is its realization, and $\RV{x}$ is a random vector and $\V{x}$ is its realization. 
Random sets and their realizations are denoted by upright sans serif and calligraphic font, respectively. For example, $\RS{X}$ is a random set and $\Set{X}$ is its realization. We denote probability density functions (pdfs) by $f(\cdot)$ and probability mass functions (pmfs) by $p(\cdot)$. 
Further, $\Set{N}(\V{x}; \V{\mu},\BM{\Sigma})$ denotes the Gaussian pdf (of random vector $\RV{x}$) with mean vector $\V{\mu}$ and covariance \vspace{0mm} matrix $\BM{\Sigma}$, $\Set{R}(x; \sigma)$ denotes the Rayleigh pdf (of scalar random variable $\rv{x}$) with scale parameter $\sigma$, $\Set{P}(n;\mu)$ denotes the Poisson pmf (of scalar random integer variable $\rv{n}$) with mean parameter $\mu$, and $\Set{B}(n;N,p)$ denotes the binomial pmf (of scalar random integer variable $\rv{n}$) of size $N$ and probability $p$. 

\vspace{2mm}
\section{Measurement Model}
\label{sec:MeasModel}
\vspace{0.5mm}

The measurement model describes the statistics of the measurements $\RV{z}_k$ conditioned on the multiobject state $\RFSst_k$ and can be characterized by the likelihood function $f(\V{z}_k|\RFSst_k)$. 
We here model the multiobject state at time $k$ by the RFS $\RFSstR_{k} \rmv= \{\stR_{k}^{(1)}\rmv,\ldots,\stR_{k}^{(n)}\}$ and the measurements by the random vector $\mevR_k = [\meR_k^{(1)}\ldots\meR_k^{(M)}]^{\T}$\rmv. 
The single-object state vectors $\stR_{k}$ are defined as $\stR_k \!=\rmv [\RV{p}_{k}^{\T} \,\, \rv{\gamma}_k]^{\T}\rmv$, where $\RV{p}_{k} = [\rv{p}_{k,1} \,\, \rv{p}_{k,2} \,\, \dot{\rv{p}}_{k,1} \,\, \dot{\rv{p}}_{k,2}]^{\T}$ is a two-dimensional (2D) position and velocity vector and $\rv{\gamma}_k\rmv$ is the object's intensity.
The modeling of $\mevR_k$ will be described in the following.

\subsection{Detector-Based Association Model}
\label{sec:AssApp}

Our modeling of $\mevR_k$ is based on $M$ data cells characterized by the random vector $\RV{c}_k = [\rv{c}_k^{(1)}\ldots \rv{c}_k^{(M)}]^{\T}$\rmv. 
Here, each $\rv{c}_k^{(m)}\rmv\rmv$, $m \rmv\in\rmv \mathcal{M} \rmv\triangleq\rmv \{1,\ldots,M\}$ represents the scalar intensity or magnitude of data cell $m$, and the cell index $m$ itself represents the corresponding positional information. We consider point objects, where the influence of one of those objects with state $\st_k \rmv\in\rmv \RFSst_{k}$ on cell $m \rmv\in\rmv \Set{M}$ is determined by an indicator function given\vspace{-2mm} by 
\vspace{0mm}
\begin{equation}
\label{eq:delta1}
\delta^{(m)}(\st_k) = 
\begin{cases}
1, & \st_k\iist \text{is in bin} \iist m \\[1mm]
0, & \st_k\iist \text{is not in bin} \iist m\ist.
\end{cases}
\vspace{0.5mm}
\end{equation}
In particular, object $\st_k$ contributes to cell $\rv{c}_k^{(m)}$ only if $\st_k$ is located in that cell, i.e., by knowing $\st_k$, we also know the index of the influenced cell.
If cell $\rv{c}_k^{(m)}$ is influenced by $\st_k$, it is distributed according to $f_{1}\big(\rv{c}^{(m)}_k\big|\st_k\big)$. 
If cell $\rv{c}_k^{(m)}$ is not influenced by any object, it is due to clutter and distributed according to $f_{0}\big(\rv{c}^{(m)}_k \big)$.

Since $\RFSstR_k$ is random, it is unknown which object contributes to which cell.
We therefore introduce the random object-cell association vector $\assvR_k \rmv\triangleq \big[\assR_k^{(1)}\ldots\assR_k^{(n)}\big]^{\T}$ with $n\rmv =\rmv |\RFSst_k|$ and entries $\assR_k^{(i)}\rmv \in\rmv \mathcal{M}$. Here, $\assR_k^{(i)} \rmv\in\rmv \mathcal{M}$ indicates that object $\st_k^{(i)}\rmv\rmv$ contributes to cell $\rv{c}_k^{(m)}$\rmv.
We assume that an object contributes to at most one cell and, conversely, that a cell is at most influenced by one object.
These assumptions result in a set of valid association vectors $\assv_k$ collected in the set $\Set{A}_k$.

We apply now a detector with threshold $\eta$ to each cell $\rv{c}^{(m)}_k\rmv\rmv$, $m \rmv\in\rmv \mathcal{M}$ resulting in ``thresholded'' measurements $\mevR_k \rmv= [\meR_k^{(1)}\ldots\meR_k^{(M)}]^{\T}$\rmv. In particular, we set $\meR_k^{(m)} \rmv= \rv{c}^{(m)}_k$ if $\rv{c}^{(m)}_k \rmv> \eta$ and $\meR_k^{(m)} \rmv= \eta$ if $\rv{c}^{(m)}_k \rmv\leq \eta$. 
In what follows, we refer to $\meR_k^{(m)} \rmv>\rmv \eta$ as detections. 
Following classical detection theory \cite{Kay98Det}, we directly obtain a statistical model for $\mevR_k$.
In particular, each detection, i.e., $\meR^{(m)}_k \rmv>\rmv \eta$, is either generated by an object or is a false alarm/clutter.
In the former case, it is distributed according to
\vspace{-1mm}
\begin{equation}
\label{eq:f_z1}
f_{1,\eta}\big(\me^{(m)}_k\big|\st_k\big) = \frac{f_{1}\big(\me^{(m)}_k\big|\st_k\big)}{\de^{(m)}(\st_k)}
\vspace{0.5mm}
\end{equation}
and, in the latter case, according to
\begin{equation}
\label{eq:f_z0}
f_{0,\eta}\big(\me^{(m)}_k\big) = \frac{f_{0}\big(\me^{(m)}_k\big)}{p^{(m)}_{\text{FA}}}.
\end{equation}
Here, the normalizations $\de^{(m)}(\st_k)$ and $p^{(m)}_{\text{FA}}$ are referred to as detection probability and false alarm probability, respectively, and can be computed as
\vspace{0mm}
\begin{align*}
\label{eq:Det1}
\de^{(m)}(\st_k) &= \int_{\eta}^{\infty} f_{1}\big(\me_k^{(m)}\big|\st_k\big)\ist\ist \mathrm{d}\me_k^{(m)} \\[1.5mm]
p^{(m)}_{\text{FA}} &= \int_{\eta}^{\infty} f_0\big(\me_k^{(m)}\big)\ist \text{d}\me_k^{(m)}. \\[-8mm]
\end{align*}

\vspace{1mm}
\subsection{The Joint Likelihood Function}
\label{sec:LikeFunc}

The likelihood function $f(\mev_k|\RFSst_k)$ 
is obtained by first computing the conditional pdf $f(\mev_k,\assv_k|\RFSst_k)$ and then marginalizing out the association variable $\assvR_k$, i.e., 
\vspace{0.5mm}
\begin{equation}
\label{eq:Like1}
f(\mev_k|\RFSst_k) \ist=\rmv \sum_{\assv_k \in\ist \mathcal{A}_{k}}  f(\mev_k,\assv_k|\RFSst_k)\ist. 
\end{equation}	
Based on the aforementioned assumptions, the conditional pdf $f(\mev_k,\assv_k|\RFSst_k)$ is found\vspace{.5mm} as
\begin{align}
f(\mev_k,\assv_k|\RFSst_k) &= \bigg( \prod_{i \ist\in\ist \Set{I}_{k}} \delta^{(\ass_k^{(i)})}\big(\st_k^{(i)}\big)\ist f\big(\me_k^{(\ass_k^{(i)})}|\st_k^{(i)}\big) \bigg) \nn \\
&\hspace{1mm}\times \prod_{m \ist\in\ist \bar{\mathcal{M}}_{\V{a}_k}} f\big(\me_k^{(m)}\big)\ist \label{eq:Like2} \\[-6mm]
\nn
\end{align}	
where we introduced $\Set{I}_k \rmv\triangleq\rmv \{1,\ldots,n\}$ and $\bar{\mathcal{M}}_{\V{a}_k} \rmv\triangleq \mathcal{M} \setminus \big\{\ass_k^{(1)}, \dots, \ass_k^{(n)} \big\}$. 
The set $\bar{\mathcal{M}}_{\V{a}_k}$ comprises all measurement indices that are not associated to any object by association~$\assv_k$.
In \eqref{eq:Like2}, the distribution $f(\me_k^{(m)}|\st_k^{(i)})$ is given by
\vspace{1.5mm} 
\begin{equation}
\label{eq:CondPDF_PI}
\hspace{-2.5mm}f\big(\me_k^{(m)}|\st_k^{(i)}\big) =
\begin{cases}
1 - p_{\text{D}}^{(m)}\big(\st_k^{(i)}\big), &\rrmv\rmv \me_k^{(m)} = \eta \\[2.5mm]
p_{\text{D}}^{(m)}\big(\st_k^{(i)}\big)\ist f_{1,\eta}\big(\me_k^{(m)}\big|\st_k^{(i)}\big), &\rrmv\rmv \me_k^{(m)} > \eta
\vspace{0.5mm} 
\end{cases} 
\end{equation}
Furthermore, the distribution $f\big(\me_k^{(m)}\big)$ reads
\vspace{0.5mm}
\begin{equation}
\label{eq:FA_Dist}
f(\me_k^{(m)})
= \begin{cases} 
1 - p_{\text{FA}}^{(m)}, & \me_k^{(m)}  =  \eta \\[2.5mm]
p_{\text{FA}}^{(m)}\ist f_{0,\eta}\big(\me_k^{(m)}\big), & \me_k^{(m)} > \eta. \\[2.5mm]
\end{cases}
\end{equation}
Our measurement model excludes the case where multiple objects are in the same cell. 
Apart from the fact that this event is rather unlikely in many applications anyway, our proposed method can still handle scenarios where some objects are in the same cell for a limited amount of time. 
This is confirmed by the very good object detection and tracking performance of our proposed method, as reported in Section~ \ref{sec:sim}.


\vspace{1mm}
\section{Multiobject Tracking with Thresholded Cell Measurements}
\label{sec:upd}
\vspace{.5mm}

In the following, we present our PMB filtering method based on the thresholded cell measurement model introduced in the previous section.

\subsection{Sequential Bayesian Estimation}

To derive our PMB filter, we use the sequential Bayesian estimation framework for RFS-based multiobject state models.
Here, the statistics of the multiobject state $\RFSstR_k$ at time $k$, conditioned on all obtained measurements $\mev_{1:k} \rmv\triangleq\rmv [\mev_1^{\T}\ldots\mev_{k}^{\T}]^{\T}$ up to time $k$, are described by the posterior pdf $f(\RFSst_k|\mev_{1:k})$. 
The estimation framework consists of a prediction and an update step.
The prediction step transforms the posterior pdf at time $k-1$, i.e., $f(\RFSst_{k-1}|\mev_{1:k-1})$, into the predicted posterior pdf at time $k$, i.e., $f(\RFSst_{k}|\mev_{1:k-1})$ using the state-transition pdf $f(\RFSst_k|\RFSst_{k-1})$.
Analogously, the update step  transforms the predicted posterior pdf into the updated posterior pdf at time $k$, i.e., $f(\RFSst_{k}|\mev_{1:k})$, using the likelihood function $f(\mev_k|\RFSst_k)$ presented in Section \ref{sec:LikeFunc} and the current measurements $\mev_k$.

Following the PMB MOT paradigm \cite{Wil:J15,Kro16,Mey18Proc,Gar18PMBM}, we here model the multiobject state $\RFSst_{k-1}$ at time $k\rmv-\rmv 1$ as the union of a Poisson RFS and an MB RFS, i.e., the posterior pdf $f(\RFSst_{k-1}|\RFSst_{k-1})$ is of PMB form.
Using standard model assumptions for $f(\RFSst_k|\RFSst_{k-1})$ \cite{Wil:J15,Kro16}, it was shown in, e.g., \cite{Wil:J15,Kro16}, that, after applying the prediction step, the predicted posterior pdf $f(\RFSst_{k}|\mev_{1:k-1})$ is again of PMB form.
In fact, $f(\RFSst_{k}|\mev_{1:k-1})$ is given according to
\vspace{1mm}
\begin{equation}
\label{eq:PredPost} 
\hspace{-1mm}f(\RFSst_k|\mev_{1:k-1})\ist = \rrmv\sum_{\RFSst_{k}^{(1)} \uplus\ist \RFSst_{k}^{(2)} =\ist\RFSst_{k}}\hspace{-3mm} f_{k|k-1}^{\text{P}}\big(\RFSst_{k}^{(1)}\big)\ist f_{k|k-1}^{\text{MB}}\big(\RFSst_{k}^{(2)}\big)\ist. 
\vspace{-0.5mm}
\end{equation} 
Here, $\sum_{\RFSst_{k}^{(1)} \uplus\ist \RFSst_{k}^{(2)} =\ist\RFSst_{k}}$ 
\vspace{-0.4mm}
denotes the sum over all disjoint decompositions of $\RFSst_{k}$ into $\RFSst_{k}^{(1)}$ and $\RFSst_{k}^{(2)}$\rmv.
Furthermore, $f_{k|k-1}^{\text{P}}\big(\RFSst_{k}\big)$ is a Poisson pdf \cite{Mah07}, which is fully parametrized by its probability hypothesis density (PHD) $\lambda_{k|k-1}(\st_k)$, and $f_{k|k-1}^{\text{MB}}\big(\RFSst_{k}\big)$ is an MB pdf \cite{Mah07} consisting of $J_{k-1}$ Bernoulli pdfs, where each Bernoulli pdf is itself parametrized by an existence probability $\ex_{k|k-1}^{(j)}$ and a spatial pdf $\sd_{k|k-1}^{(j)}(\st)$, $j \rmv\in \Set{J}_{k-1} \triangleq \{1,\ldots,J_{k-1}\}$.
See, e.g., \cite{Mah07}, for more information on RFSs and their statistical description.
In the following two subsections, we present the exact and approximate update steps.

\vspace{1mm}
\subsection{Exact Update Step}
\label{sec:ExUpd}
\vspace{0.5mm}

Note that the current measurements $\mev_k$ have already been observed and are therefore known. 
It is therefore also known which data cells $m \rmv\in\rmv \Set{M}$ have an intensity above threshold $\eta$ and which do not.
We here introduce the index set $\Set{M}_k^{\text{D}} \subseteq \Set{M}$ of all detections, i.e., all cell indexes with $\me_k^{(m)} > \eta$ and the index set $\Set{M}_k^{\text{M}} = \Set{M} \setminus \Set{M}_k^{\text{D}}$ of all missed detections, i.e., all cell indexes with $\me_k^{(m)} \rmv= \eta$.

After applying the update step, as we show in Appendix \ref{sec:App_A}, the posterior pdf $f(\RFSst_{k}|\mev_{1:k})$ is no longer a PMB pdf but a Poisson/MB mixture (PMBM) pdf,
which is given by
\vspace{0.5mm}
\begin{equation}
\label{eq:upd1}
f(\RFSst_{k}|\mev_{1:k}) \ist=\! \sum_{\RFSst_{k}^{(1)} \uplus\ist \RFSst_{k}^{(2)} =\ist\RFSst_{k}}\hspace{-3mm} f^{\text{P}}\big(\RFSst_{k}^{(1)}\big)\ist f^{\text{MBM}}\big(\RFSst_{k}^{(2)}\big)\ist. 
\vspace{-0.5mm}
\end{equation}
Here, $f^{\text{P}}\big(\RFSst_{k}\big)$ is a Poisson pdf and $f^{\text{MBM}}\big(\RFSst_{k}\big)$ is a multi-Bernoulli mixture (MBM) pdf; expressions of both will be presented now.

The Poisson pdf $f^{\text{P}}\big(\RFSst_{k}\big)$ in \eqref{eq:upd1} is fully characterized by the posterior PHD $\lambda(\st_k)$, which is given by
\begin{align}
\lambda(\st_k) = \bigg(\sum_{m\ist\in\ist\Set{M}^{\text{M}}_k}\delta^{(m)}(\st_k)\ist \frac{1-p_{\text{D}}^{(m)}(\st_k)}{1-p_{\text{FA}}^{(m)}}\ist\bigg) \lambda_{k|k-1}(\st_k)\ist. \nn \\[-3mm]
\label{eq:upd_Pois} \\[-7mm]
\nn
\end{align}
Here, $p_{\text{D}}^{(m)}(\st_k)$ and $p_{\text{FA}}^{(m)}$ are the detection probability and false alarm probability, respectively, defined in Section \ref{sec:AssApp} and $\lambda_{k|k-1}(\st_k)$ is the predicted posterior PHD characterizing $f_{k|k-1}^{\text{P}}\big(\RFSst_{k}\big)$ in \eqref{eq:PredPost}. 

In order to describe $f^{\text{MBM}}\big(\RFSst_{k}\big)$ in \eqref{eq:upd1}, 
\vspace{-0.5mm}
we first introduce the random object-cell association vector $\assvR'_k\triangleq [\assR_k^{\prime(1)}\ldots\assR_k^{\prime(J_{k})}]^{\T}$ where $J_k = J_{k-1}+M^{\text{D}}_k$ and $M_k^{\text{D}} \triangleq |\Set{M}_k^{\text{D}}|$. 
Whereas the association vector $\assvR_k$ defined in Section \ref{sec:AssApp} describes the association between objects and data cells, the vector $\assvR'_k$ describes now the association of potential objects, i.e., Bernoulli components, with data cells.  
More precisely, $\assvR'_k$ has entries $\assR_k^{\prime(j)} \rmv\in \rmv\{0\}\rmv \cup\rmv \Set{M}$ for $j \rrmv\in\rrmv \Set{J}_{k-1}$ and $\assR_k^{\prime(j)} \rmv\in\rmv \{0,1\}$ for $j \rrmv\in\rmv \Set{J}_k \setminus \Set{J}_{k-1}$.
Here, for $j \rrmv\in\rmv \Set{J}_{k-1}$, $\assR_k^{\prime(j)} \rmv= 0$ indicates that the object $\stR_k$ modeled by Bernoulli component $j$ does not exist, $\assR_k^{\prime(j)} \rmv\in \Set{M}^{\text{M}}_k$ that it exists but did not generate a detection in cell $m\in\Set{M}^{\text{M}}_k$, and $\assR_k^{\prime(j)} \rmv\in\rmv\Set{M}^{\text{D}}_k$ that it exists and generated detection $m \in \Set{M}_k^{\text{D}}$. 
Furthermore, for $j \rmv\in\rmv \Set{J}_k \setminus \Set{J}_{k-1}$, $\assR_k^{\prime(j)} \rmv= 1$ indicates that detection $m \rmv\in \Set{M}_k^{\text{D}}$ was generated either by an object $\stR_k$ modeled by the Poisson pdf or by clutter and, conversely, $\assR_k^{\prime(j)} \rmv=\rmv 0$ that detection $m \rmv\in\rmv \Set{M}_k^{\text{D}}$ was neither generated by an object modeled by the Poisson RFS nor by clutter. 
Here, the mapping between $j\in\Set{J}_k \rmv\setminus\rmv \Set{J}_{k-1}$ and $m\rmv\in\rmv\Set{M}_k^{\text{D}}$ is arbitrary but unique.
Note that there is exactly one $j \rmv\in\rmv \Set{J}_k \setminus \Set{J}_{k-1}$, i.e., a new Bernoulli component, for each detection $m\rmv\in\rmv\Set{M}_k^{\text{D}}$. 
All valid association vectors $\assv'_k$ form the set $\mathcal{A}'_{k}$. 

\vspace{1mm}
Using $\assvR'_k$, we can now express $f^{\text{MBM}}\big(\RFSst_{k}\big)$ according to
\vspace{-0.5mm}
\begin{equation}
\label{eq:upd_MB}
f^{\text{MBM}}\big(\RFSst_{k}\big) = \sum_{\assv'_k\ist\in\ist\mathcal{A}'_{k}} \ist p(\assv'_k)\ist\ist f^{\text{MB}}_{\assv'_k}(\RFSst_k)\ist. 
\vspace{-1mm}
\end{equation}
Note that there is exactly one mixture component for each valid association vector $\assv'$ and that the corresponding weight is given by the association probability $p(\assv'_k)$, which in turn is given by
\begin{equation}
\label{eq:upd4}
p(\assv'_k) \propto \prod_{j\in\Set{J}_k}\ist \assw_k^{(j,\ass_k^{\prime(j)})}
\vspace{0mm}
\end{equation}
for $\assv'_k \rmv\in\rmv \mathcal{A}'_{k}$\ist, and by $p(\assv'_k) \rmv= 0$ for $\assv'_k \rmv\notin\rmv \mathcal{A}'_{k}$.
Furthermore, the MB pdfs $f^{\text{MB}}_{\assv'_k}(\RFSst_k)$ consist of $J_k = J_{k-1} + M^{\text{D}}_k$ Bernoulli components, where each of them is parametrized by an existence probability $\ex_k^{(j,m)}$ and a spatial pdf $\sd^{(j,m)}(\st_k)$.
In the following, we provide expressions for $\ex_k^{(j,m)}$\rmv, $\sd^{(j,m)}(\st_k)$, and $\assw_k^{(j,m)}$.

In fact, for Bernoulli component $j \rmv\in\rmv\Set{J}_{k-1}$ and $m\rmv = 0$\ist, we have
$\ex_k^{(j,0)} \rmv= 0$\ist, $\sd^{(j,0)}(\st_k)$ undefined, and 
\vspace{1mm}
\begin{align}
\assw_k^{(j,0)} &= 1 - \ex_{k|k-1}^{(j)}\ist.  \label{eq:upd_para1_3} \\[-5mm]
\nn
\end{align} 
Here, $\ex_k^{(j,0)}\rrmv\rmv =\rrmv 0$ indicates that the object $\stR_k$ modeled by Bernoulli component $j$ does not exist; $\sd^{(j,0)}(\st_k)$ remains thus undefined. The likelihood of this event is given by \eqref{eq:upd_para1_3}. 

Next, for $j \rmv\in\rmv\Set{J}_{k-1}$ and $m \in \Set{M}^{\text{M}}_k$, we have $\ex_k^{(j,m)} \rmv= 1$ and
\vspace{0mm}
\begin{align}
\sd^{(j,m)}(\st_k) & \ist=\ist \frac{\delta^{(m)}(\st_k)\ist \big(1\rmv-\rmv p_{\text{D}}^{(m)}(\st_k)\big) \ist \sd_{k|k-1}^{(j)}(\st_k)}{b_k^{(j,m)}} \label{eq:upd_para2_2}  \\[2mm]
\assw_k^{(j,m)} &\ist=\ist \frac{\ex_{k|k-1}^{(j)}\ist b_k^{(j,m)}}{1-p_{\text{FA}}^{(m)}} \label{eq:upd_para2_3}  \\[-4mm]
\nn
\end{align}
with $b_k^{(j,m)}\rrmv\triangleq\rmv \int\rmv\delta^{(m)}(\st_k) \big(1 - p_{\text{D}}^{(m)}(\st_k)\big)\ist\sd_{k|k-1}^{(j)}(\st_k)\ist \text{d}\st_k$. 
\vspace{-0.5mm}
Here, $\ex_k^{(j,m)} \rmv= 1$ indicates that the object $\stR_k$ modeled by Bernoulli component $j$ exists but was not detected in cell $m$. Its spatial pdf is distributed according to \eqref{eq:upd_para2_2}, and the likelihood of this event is given by \eqref{eq:upd_para2_3}. 

Next, for $j \rmv\in\rmv\Set{J}_{k-1}$ and $m \in \mathcal{M}_k^{\text{D}}$, we have $\ex_k^{(j,m)}\rmv = 1$, and 
\vspace{-0.5mm}
\begin{align}
\sd^{(j,m)}(\st_k) & \ist=\ist \frac{\delta^{(m)}(\st_k)\ist p_{\text{D}}^{(m)}(\st_k)\ist f_{1,\eta}(\me_k^{(m)}|\st_k) \ist \sd_{k|k-1}^{(j)}(\st_k)}{c_k^{(j,m)}} \nn \\[-2.5mm]
\label{eq:upd_para3_2} \\[2mm]
\assw_k^{(j,m)} &\ist=\ist \ex_{k|k-1}^{(j)}\ist c_k^{(j,m)} \label{eq:upd_para3_3}
  \\[-3mm]
\nn
\end{align} 
with 
\vspace{-0.6mm}
$c_k^{(j,m)}\rrmv\triangleq\rmv\int\rmv\delta^{(m)}(\st_k)\ist p_{\text{D}}^{(m)}(\st_k)\ist f_{1,\eta}(\me_k^{(m)}|\st_k)\ist\sd_{k|k-1}^{(j)}(\st_k)$ $\times\text{d}\st_k$.
Here, $\ex_k^{(j,m)} \rmv= 1$ indicates that the object $\stR_k$ modeled by Bernoulli component $j$ exists and was detected in cell $m$. Its spatial pdf is distributed according to \eqref{eq:upd_para3_2}, and the likelihood of this event is given by \eqref{eq:upd_para3_3}.

There are also new Bernoulli components  $j \rmv\in\rmv \Set{J}_{k}\setminus\Set{J}_{k-1}$ and $m \rmv\in\rmv \Set{M}_k^{\text{D}}$, where we have 
\vspace{1.5mm}
\begin{align}
\ex_k^{(j,1)} & \ist=\ist \frac{d_k^{(j)}}{p_{\text{FA}}^{(m)}\ist f_{0,\eta}(\me_k^{(m)}) + d_k^{(j)}} \label{eq:upd_para4_1} \\[4mm]
\sd^{(j,1)}(\st_k) & \ist=\ist \frac{\delta^{(m)}(\st_k)\ist p_{\text{D}}^{(m)}(\st_k)\ist f_{1,\eta}(\me_k^{(m)}|\st_k)\ist \lambda_{k|k-1}(\st_k)}{d_k^{(j)}} \nn \\[-2.5mm]
\label{eq:upd_para4_2} \\[2.5mm]
\assw_k^{(j,1)} &\ist=\ist p_{\text{FA}}^{(m)}\ist f_{0,\eta}(\me_k^{(m)}) + d_k^{(j)} \label{eq:upd_para4_3} \\[-2mm]
\nn
\end{align}
with 
\vspace{-0.5mm}
$d_k^{(j)} \rrmv\triangleq\rmv \int\rmv \delta^{(m)}(\st_k)\ist p_{\text{D}}^{(m)}(\st_k)\ist f_{1,\eta}(\me_k^{(m)}|\st_k)\ist\lambda_{k|k-1}(\st_k)\text{d}\st_k$.
Here, $\ex_k^{(j,1)}$ in \eqref{eq:upd_para4_1} is the probability that detection $m$ was generated by object $\stR_k$ modeled by the Poisson RFS and not due to clutter.  
The state of this object is distributed according to \eqref{eq:upd_para4_2}, and the likelihood of the events that detection $m$ was generated either by an object modeled by the Poisson RFS or by clutter is given by \eqref{eq:upd_para4_3}.
Finally, we have $\ex_k^{(j,0)} \rmv= 0$, $\sd^{(j,0)}(\st_k)$ undefined, and\vspace{-2mm} $\assw_k^{(j,0)} \rmv= 1$.

\subsection{Approximate Update Step}
\label{sec:AppUpd}

We now approximate the MBM pdf in \eqref{eq:upd_MB} by an MB pdf.
We start by extending the association alphabet $\mathcal{A}'_{k}$ to $\mathcal{A}''_{k} \triangleq (\{0\}\cup\Set{M})^{J_{k-1}}\times\{0,1\}^{M_k^{\text{D}}}$. 
Since $p(\assv'_k)\rmv =\rmv 0$ for $\assv'_k \rmv\notin\rmv \mathcal{A}'_{k}$, this extension does not change the association pmf.  Next, we approximate $p(\assv'_k)$ according to 
\vspace{3mm}
\begin{equation}
\label{eq:appox1}
p(\assv'_k) \approx \prod_{j \in \Set{J}_k}\ist p\big(\ass_k^{\prime(j)}\big) \ist, \quad \assv'_k \rmv\in\rmv \mathcal{A}''_{k}
\vspace{-1mm}
\end{equation}
with 
\vspace{1.5mm}
\[
p\big(\ass_k^{\prime(j)}\big) =\rmv \sum_{\sim\assv_k^{\prime (j)}} \ist p(\assv'_k) \ist.
\vspace{0mm}
\]	
Here, $\sim \rmv \assv_k^{\prime (j)}$ denotes the vector of all $a_k^{\prime(j')}$ with $j' \rmv\in\rmv \Set{J}_k \rmv\setminus j$.
Note that a fast and scalable approximate solutions for $p\big(\ass_k^{\prime(j)}\big)$ is provided by the SPA \cite{Wil14,Wil:J15,Mey18Proc}.

As was shown in, e.g., \cite{Kro21TBD}, by inserting approximation \eqref{eq:appox1} into \eqref{eq:upd_MB}, the MBM pdf simplifies to an MB pdf.
The corresponding existence probabilities and spatial pdfs for \linebreak $j \rmv\in\rmv \Set{J}_{k-1}$ are given by
\vspace{2mm}
\begin{align}
\ex_k^{(j)} &=  \sum_{\ass_k^{\prime(j)} \in\ist \mathcal{M}} p(\ass_k^{\prime(j)})  \label{eq:approx4} \\[2.5mm]
\sd^{(j)}(\st_k) &= \frac{1}{\ex_k^{(j)}} \rmv\rmv \sum_{\ass_k^{\prime(j)} \in\ist \mathcal{M}} p(\ass_k^{\prime(j)}) \ist \sd^{(j,\ass_k^{\prime(j)})}(\st_k)\label{eq:approx5} \\[-4.5mm]
\nn
\end{align}
and for $j \rmv\in\rmv \Set{J}_{k} \setminus \Set{J}_{k-1}$ by
\vspace{1.2mm}
\begin{align}
\ex_k^{(j)} &=  p(\ass_k^{\prime(j)} = 1)\iist \ex_k^{(j,1)} \label{eq:approx6} \\[3mm]
\sd^{(j)}(\st_k) &= \ist \sd^{(j,1)}(\st_k)\ist. \label{eq:approx7} \\[-4.5mm]
\nn
\end{align}

Note that according to \eqref{eq:approx6} and \eqref{eq:approx7}, a new Bernoulli component is generated for each detection in each update step.
This results in a linear increase in the number of Bernoulli components over time. In order to counteract this increase and limit the number of Bernoulli components, we employ the concept of recycling, where Bernoulli components with low existence probabilities are transferred to the Poisson part of the multiobject state RFS.
More precisely, we transfer all Bernoulli components $j \rmv\in\rmv \mathcal{J}_k^{\text{R}} \rmv\subseteq\rmv \Set{J}_k$ with an existence probability $\ex_k^{(j)}$ below the threshold $\eta_{\text{R}}$. 
This yields the approximated posterior PHD 
\vspace{2mm}
\begin{equation}
\label{eq:approx8}
\tilde{\PHD}(\st_k) = \PHD(\st_k) + \sum_{j\ist\in\ist \mathcal{J}_k^{\text{R}}} \ex_k^{(j)}\ist \sd^{(j)}(\st_k) \nn
\end{equation}
where $\PHD(\st_k)$ is given by \eqref{eq:upd_Pois}, $\ex_k^{(j)}$ is given by \eqref{eq:approx4} or \eqref{eq:approx6} and $\sd^{(j)}(\st_k)$ by \eqref{eq:approx5} or \eqref{eq:approx7}, respectively. 
In contrast to pruning Bernoulli components with low existence probabilities, recycling does not discard any information.

\section{Multiobject Tracking with Point Measurements}
\label{sec:PMM_wAI}
\vspace{1mm}

In the following, we derive a likelihood function for point measurements that also uses AM.
Contrary to our cell measurement model where the measurements are described by the random vector $\mevR_k = [\meR_k^{(1)}\ldots\meR_k^{(M)}]^{\T}$ and where the measurement index represents the positional information, point measurements are typically modeled by an unordered set $\RS{Z}_k = \{\meR_k^{(1)},\ldots,\meR_k^{(\rv{M}_k)}\}$. Here, positional information is modeled explicitly by position-related measurements and not by the measurement index.

\subsection{Explicit Location Information and Clutter Cardinality}
\vspace{1mm}

Let $\big[\meR^{(m)}_{k,1}\, \meR^{(m)}_{k,2}\big]^{\T}$ be the position-related measurement, e.g., in range-bearing or in Cartesian coordinates, corresponding to amplitude measurement $\meR^{(m)}_k$\rmv, and let $\RV{y}_k^{(m)} = \big[\meR^{(m)}_{k} \ist\ist\ist \meR^{(m)}_{k,1} \ist\ist\ist \meR^{(m)}_{k,2}\big]^{\T} $ be the corresponding joint measurement vector. 
Position-related measurements can be obtained by breaking up the cell structure and converting cell indices to position-related measurements.
A joint measurement vector $\RV{y}_k^{(m)}\rmv\rmv$, can be statistically described by the following joint pdfs
\vspace{-2mm}
\begin{align}
f_1(\V{y}_k^{(m)}|\st_k) &= f_{1,\text{p}}(\me_{k,1}^{(m)}\rmv, \me_{k,2}^{(m)}|\st_k)\ist f_{1,\eta}(\me_k^{(m)}|\st_k) \label{eq:f1} \\[2.5mm]
f_0(\V{y}_k^{(m)})&= f_{0,\text{p}}(\me_{k,1}^{(m)}\rmv, \me_{k,2}^{(m)})\ist f_{0,\eta}(\me_k^{(m)}). \label{eq:f0} \\[-4.5mm]
\nn
\end{align}
Here, $f_{1,\eta}(\me_k^{(m)}|\st_k)$ and $f_{0,\eta}(\me_k^{(m)})$ are given by \eqref{eq:f_z1} and \eqref{eq:f_z0}, respectively, and $f_{1,\text{p}}(\me_{k,1}^{(m)}, \me_{k,2}^{(m)}|\st_k)$ and $f_{0,\text{p}}(\me_{k,1}^{(m)}, \me_{k,2}^{(m)})$ are the pdfs of the position-related measurement for the cases that measurement $\RV{y}^{(m)}_k$ is generated by object $\st_k$ or by false alarm/clutter, respectively.
Given these basic definitions, we apply now a series of manipulations and approximations in order to transform the likelihood function for thresholded cell measurements in \eqref{eq:Like1} and \eqref{eq:Like2} into a likelihood function for point measurements that also uses AM.

We start by noting that in the likelihood function for thresholded cell measurements the information about the number of clutter measurements
\vspace{-0.8mm} 
is contained in the term $\prod_{\bar{\Set{M}}_{\assv_k}} f(\me_k^{(m)})$ with $f(\me_k^{(m)})$ given by \eqref{eq:FA_Dist}.
Here $f(\me_k^{(m)})$ can be rewritten as follows: 
We first explicitly represent the events $\meR_k^{(m)} \rmv=\rmv \eta$ and $\meR_k^{(m)} \rmv>\rmv \eta$ by the Bernoulli random variable $\rv{b}_k^{(m)} \rmv\in\rmv \{0,1\}$, where $\rv{b}_k^{(m)} \rmv=\rmv 0$ models the event that $\meR_k\rmv=\rmv\eta$ and $\rv{b}_k^{(m)}\rmv =\rmv 1$ the event that $\meR_k^{(m)}\rmv>\rmv\eta$. Using $\rv{b}_k^{(m)}$\rmv, we can now rewrite $f(\me_k^{(m)})$ in \eqref{eq:FA_Dist} according to 
\[
f\big(\me_k^{(m)}\rmv,b_k^{(m)}\big) = f\big(\me_k^{(m)}|b_k^{(m)}\big)\ist p\big(b_k^{(m)}\big)  
\]
with
\begin{equation}
\label{eq:prior_b}
p\big(b_k^{(m)}\big)
= \begin{cases} 
1 - p_{\text{FA}}^{(m)}\rmv, & b_k^{(m)} = 0 \\[2.5mm]
p_{\text{FA}}^{(m)}\rmv, & b_k^{(m)} = 1\ist. 
\end{cases}
\end{equation}
Furthermore, $f\big(\me_k^{(m)}|b_k^{(m)} \rrmv=\rrmv 1\big) = f_{0,\eta}\big(\me_k^{(m)}\big)$ and, since $f(\me_k^{(m)}|b_k^{(m)} \rmv=\rmv 0)$ is meaningless, it is not defined.

In point measurement models, the information about the number of clutter measurements given by \eqref{eq:prior_b} is typically represented by a single cardinality variable $\rv{M}_k^{\text{FA}}$.
In order to derive a likelihood function for point measurements, we therefore define 
\vspace{0mm}
\begin{equation}
\label{eq:FA_Meas1}
\rv{M}_k^{\text{FA}} \ist\ist\triangleq\rmv\rmv \sum_{m \ist\in\ist \bar{\Set{M}}_{\assv_k}} \rv{b}_k^{(m)}. 
\vspace{-2mm}
\end{equation} 
Since $\rv{b}_k^{(m)}$ is Bernoulli distributed according to \eqref{eq:prior_b}, and given the fact that the sum of Bernoulli distributed random variables is Binomial distributed \cite{Pap01}, $\rv{M}_k^{\text{FA}}$ in \eqref{eq:FA_Meas1} is binomial distributed too, i.e.,
\vspace{0.5mm}
\begin{align}
\label{eq:FA_Meas2}
p(M_k^{\text{FA}}) &=\ist \Set{B}\big(M_k^{\text{FA}};|\bar{\Set{M}}_{\assv_k}|,\ist p_{\text{FA}}\big) \nn \\[1mm]
&\triangleq \binom{|\bar{\Set{M}}_{\assv_k}|}{M_k^{\text{FA}}} (p_{\text{FA}})^{M_k^{\text{FA}}} (1-p_{\text{FA}})^{|\bar{\Set{M}}_{\assv_k}|-M_k^{\text{FA}}} 
\vspace{-0.5mm}
\end{align}
where $\binom{|\bar{\Set{M}}_{\assv_k}|}{M_k^{\text{FA}}}$ is the binomial coefficient of $|\bar{\Set{M}}_{\assv_k}|$ and $M_k^{\text{FA}}$\rmv.

For later use, we approximate the binomial pmf $p(M_k^{\text{FA}})$ in \eqref{eq:FA_Meas2} by a Poisson pmf, i.e.,
\vspace{0mm}
\begin{equation}
\label{eq:Approx_FA}
p(M_k^{\text{FA}}) \approx\ist \Set{P}\big(M_k^{\text{FA}};\mu_{\text{FA}}\big) \triangleq \frac{(\mu_{\text{FA}})^{M_k^{\text{FA}}} e^{-\mu_{\text{FA}}}}{M_k^{\text{FA}}!} 
\vspace{0mm}
\end{equation}
with $\mu_{\text{FA}} \rmv=\rmv p_{\text{FA}}\ist |\bar{\Set{M}}_{\assv_k}| \approx p_{\text{FA}}\ist M$\rmv.
According to the Poisson limit theorem \cite{Pap01}, the approximation error of \eqref{eq:Approx_FA} is small if both $p_{\text{FA}}$ is small and $M$ is large.

\subsection{Object-Measurement Associations and Likelihood Function}

Next, we note that conventional point measurement models allow any possible object-measurement association, theoretically, with non-zero probability.
We now define the random object-measurement association vector for point measurements as $\overline{\assvR}_k \rmv=\rmv [\overline{\assR}_k^{(1)}\ldots\overline{\assR}_k^{(n)}]^{\T}$ with entries $\overline{\assR}_k^{(i)} \rrmv\in\rmv \{0\} \rmv\cup\rmv \Set{M}_k$, $\Set{M}_k \triangleq \{1,\ldots,M_k\}$, and where $M_k$ is the number of point measurements $\RV{y}_k^{(m)}$\rmv. 
Here, $\overline{\assR}_k^{(i)} \rmv=\rmv 0$ indicates that object $\st_k^{(i)}$ did not generate any measurement and $\overline{\assR}_k^{(i)} \rmv= m  \rmv\in\rmv\Set{M}_k$ that it generated measurement $\RV{y}_k^{(m)}$\rmv.
All valid association vectors $\overline{\assv}_k$ are collected in the set $\overline{\Set{A}}_k$.

We now model all $\overline{\assv}_k$ as equally likely, i.e., 
\begin{equation}
\label{eq:PriorAss}
p(\overline{\assv}_k|\RFSst_k) = \frac{1}{N_{\overline{\assv}_k}}
\end{equation}
where $N_{\overline{\assv}_k} \rmv=\rmv |\overline{\Set{A}}_k|$ is the number of association vectors.  
Here, $N_{\overline{\assv}_k}$ can be determined by using basic results from combinatorics.
In fact, $N_{\overline{\assv}_k}$ is equivalent to the number of possible combinations of drawing $M_k^{\text{D}}$ object detections out of $M_k$ measurements, where the draws are without replacement and with the drawing order respected. This leads to 
\vspace{0mm}
\begin{equation}
\label{eq:NumAss}
N_{\overline{\assv}_k} = \frac{M_k!}{(M_k-M_k^{\text{D}})!} = \frac{M_k!}{M_k^{\text{FA}}!}\ist.
\vspace{1mm}
\end{equation}

Putting all together, we find the joint pdf of $\RV{y}_k$, $\overline{\assvR}_k$, and $\rv{M}_k^{\text{FA}}$, conditioned on $\RFSst_k$, according to
\begin{align}
\hspace{-4mm}f(\V{y}_k,M_k^{\text{FA}}\rmv,\overline{\assv}_k|\RFSst_k) &=\ist \frac{1}{N_{\overline{\assv}_k}}\Big(\prod_{i \ist\in\ist \Set{I}_{k}} \ist f\big(\V{y}_k^{(\overline{\ass}_k^{(i)})}|\st_k^{(i)}\big) \Big) \nn\\
&\hspace{1mm}\times\Big(\prod_{m\ist\in\ist\bar{\Set{M}}_{\overline{\assv}_k}} f_0(\V{y}_k^{(m)})\Big) \iist p(M_k^{\text{FA}}) \label{eq:Point1} \\[-6mm]
\nn
\end{align}
where $\bar{\Set{M}}_{\overline{\assv}_k} \triangleq \Set{M}_k \setminus \{\overline{\ass}_k^{(1)},\ldots,\overline{\ass}_k^{(n)}\}$.
In line with \eqref{eq:CondPDF_PI}, $f\big(\V{y}_k^{(\overline{\ass}_k^{(i)})}|\st_k^{(i)}\big)$ is given by 
\vspace{1.5mm}
\[
\label{eq:Point2}
f\big(\V{y}_k^{(\overline{\ass}_k^{(i)})}|\st_k^{(i)}\big) = 
\begin{cases}
1 - p_{\text{D}}\big(\st_k^{(i)}\big), &\rrmv \overline{\ass}_k^{(i)} \rmv= 0 \\[2.5mm]
p_{\text{D}}\big(\st_k^{(i)}\big) f_1(\V{y}_k^{(m)}|\st_k), &\rrmv \overline{\ass}_k^{(i)} \rmv\in \Set{M}_k. 
\end{cases}
\vspace{1mm}
\]
Here, $f_1(\V{y}_k^{(m)}|\st_k)$ is given by \eqref{eq:f1}, $f_0(\V{y}_k^{(m)})$ and $p(M_k^{\text{FA}})$ by \eqref{eq:f0} and \eqref{eq:FA_Meas2}, respectively.

By finally inserting $N_{\overline{\assv}_k}$ given by \eqref{eq:NumAss} and the Poisson approximation of $p(M_k^{\text{FA}})$ given by \eqref{eq:Approx_FA} into \eqref{eq:Point1}, we get the approximate joint distribution $\tilde{f}(\V{y}_k,M_k^{\text{FA}},\overline{\assv}_k|\RFSst_k)$. Due to the relation $M_k = M_k^{\text{D}} + M_k^{\text{FA}}$, $\tilde{f}(\V{y}_k,M_k^{\text{FA}},\overline{\assv}_k|\RFSst_k)$ is equivalent to $\tilde{f}(\V{y}_k,M_k,\overline{\assv}_k|\RFSst_k)$, given by
\vspace{0.5mm}
\begin{align}
\tilde{f}(\V{y}_k,M_k,\overline{\assv}_k|\RFSst_k) &= \frac{e^{-\mu_{\text{FA}}}}{M_k!} \Big(\prod_{i \ist\in\ist \Set{I}_{k}} \ist f\big(\V{y}_k^{(\overline{\ass}_k^{(i)})}\rmv,\overline{\ass}_k^{(i)}|\st_k^{(i)}\big) \Big) \nn \\[1mm]
&\hspace{2mm}\times\prod_{m \ist\in\ist \bar{\mathcal{M}}_{\overline{\V{a}}_k}}\rrmv \lambda_{\text{FA}}\big(\V{y}^{(m)}_k\big) \label{eq:meas2_5_3} \\[-6mm]
\nn
\end{align}
with $\lambda_{\text{FA}}(\V{y}_k) \rmv\triangleq\rmv \mu_{\text{FA}}\ist f_0(\V{y}_k)$.
\vspace{0.2mm} 
By finally marginalizing out $\overline{\assv}_k$, we find the likelihood function for point measurements as 
\vspace{0.5mm}
\begin{equation}
\label{eq:Like_PM}
f(\V{y}_k,M_k|\RFSst_k) \ist=\rmv \sum_{\overline{\assv}_k \in\ist \overline{\mathcal{A}}_{k}}\ist  \tilde{f}(\V{y}_k,M_k,\overline{\assv}_k|\RFSst_k)\ist. 
\vspace{0mm}
\end{equation}
Note that here, $\RV{y}_k$ is a random vector consisting of $\rv{M}_k$ subvectors $\RV{y}_k^{(m)}$\rmv. 
However, in RFS-based MOT with point measurements, the measurements are typically represented as a set $\Set{Y}_k = \{\RV{y}_k^{(1)},\ldots,\RV{y}_k^{(M_k)}\}$. The corresponding pdf is given by $f(\Set{Y}_k|\RFSst_k) = M_k!\ist f(\V{y}_k,M_k|\RFSst_k)$ \cite{Mah07}. 
If the AM information $\meR^{(m)}_{k}$ is removed from $\RV{y}_k^{(m)}$\rmv, we obtain the corresponding point measurement model without AM. 
Based on \eqref{eq:meas2_5_3} and \eqref{eq:Like_PM}, a PMB filter using point measurements with AM can be derived analogously as in \cite{Wil:J15}.
\vspace{1mm}

\section{Simulation Study}
\vspace{0mm}
\label{sec:sim}

In the following, we perform a simulation study in which we compare the performance of our proposed PMB filter with thresholded pixel measurements (PMB-CM) with that of a PMB filter that uses point measurements with (PMB-AM) and without (PMB) AM. But first, we start with a short description of the underlying simulation setup.

\subsection{Simulation Setup}

We consider a 2D simulation scenario with a region of interest (ROI) of $[\text{0}\text{m},\text{32}\text{m}] \times [\text{0}\text{m},\text{32}\text{m}]$. 
We simulated 10 objects during 200 time steps. 
Recap that the single-object state vectors are defined as $\stR_k \!=\rmv [\RV{p}_{k}^{\T} \,\, \rv{\gamma}_k]^{\T}\rmv$, where $\RV{p}_{k} = [\rv{p}_{k,1} \,\, \rv{p}_{k,2} \,\, \dot{\rv{p}}_{k,1} \,\, \dot{\rv{p}}_{k,2}]^{\T}$ is the object's 2D position and 2D velocity and $\rv{\gamma}_k$ the object's intensity (cf. Section \ref{sec:MeasModel}). 
The objects appear at various times before time step 30 at randomly chosen positions in the ROI, and they disappear at various times after time step 170 or when they leave the ROI.
The objects' initial velocities are drawn from $f_{\text{v}}(\dot{p}_{k,1}, \, \dot{p}_{k,2}) = \Set{N}(\dot{p}_{k,1},\, \dot{p}_{k,2};\V{0}_2,\sigma^{2}_{\text{v}}\V{I}_2)$ with $\sigma^{2}_{\text{v}} \rmv=\rmv 10^{-2}$, and the objects' initial intensities are set to $10$.

The object states' evolution is modeled as follows: 
The kinematic part $\RV{p}_{k}$ evolves according to the nearly constant velocity motion model, which is given by $\RV{p}_k \rmv= \M{A}\ist \RV{p}_{k-1} + \M{W}\ist \RV{w}_k$ with $\M{A} \rmv\in\rmv \mathbb{R}^{4\times 4}$ and $\M{W} \rmv\in\rmv \mathbb{R}^{4\times 2}$ chosen as in \cite{Kro16}. The driving noise $\RV{w}_k$ is distributed according to $\Set{N}(\V{w}_{k};\V{0}_2,10^{-3}\M{I}_2)$. The object intensities remain constant over time.

The measurement consists of $32\times32$ data cells of square size, each of which has a side length of $1$m. The data is generated by modeling the cell intensities as Rayleigh distributed, i.e., $f_{1}\big(c_k^{(m)}\big|\st_k\big) = \Set{R}(c^{(m)}_k; \sqrt{\gamma_k + \sigma^{2}_{\text{n}}} \ist )\vspace{-.5mm}$ and $f_{0}(y_k^{(m)}) = \Set{R}\big(y^{(m)}_k; \sigma_{\text{n}}\big)$ and drawing samples from them. In radar applications, this model is referred to as Swerling~ 1~ \cite{Ris19}. 
We set $\sigma^{2}_{\text{n}} \rmv=\rmv 1$ leading to a signal-to-noise ratio (SNR) of $10$. 
If there are several objects in the same cell, lets say $\st^{(m_1)}_k\rmv, \dots, \st^{(m_I)}_k$\rmv, then 
\vspace{0.3mm}
the cell intensity $\rv{c}^{(m)}_k$ is sampled from $\Set{R}\Big(c^{(m)}_k; \sqrt{\gamma^{(m_1)}_k + \dots + \gamma^{(m_I)}_k + \sigma^{2}_{\text{n}}} \Big) $. 
\vspace{-0.2mm}
In PMB-CM, we set 
$f_{1}\big(c_k^{(m)}\big|\st_k\big) = \Set{R}(c^{(m)}_k; \sqrt{\gamma_k + \sigma^{2}_{\text{n}}} \ist )$ 
\vspace{0.2mm}
and $f_{0}(y_k^{(m)}) = \Set{R}\big(y^{(m)}_k; \sigma_{\text{n}}\big)$. 
As discussed in Section \ref{sec:AssApp}, 
\vspace{-0.2mm}
for a given threshold $\eta$, this directly induces $f_{1,\eta}\big(z_k^{(m)}\big|\st_k\big)$ and  $f_{0,\eta}\big(z_k^{(m)}\big)$.

For PMB and PMB-AM, position-related measurement information is represented explicitly according to $\big[\meR^{(m)}_{k,1}\, \meR^{(m)}_{k,2}\big]^{\T}$\rmv\rmv, which necessitates the further specification of $f_{1,\text{p}}(\me_{k,1}^{(m)}\rmv, \me_{k,2}^{(m)}|\st_k)$ in \eqref{eq:f1} and $f_{0,\text{p}}(\me_{k,1}^{(m)}\rmv, \me_{k,2}^{(m)})$ in \eqref{eq:f0}.
We follow the frequently used convention for point measurement filtering and model $f_{1,\text{p}}(\me_{k,1}^{(m)}\rmv, \me_{k,2}^{(m)}|\st_k)$ as a Gaussian pdf $\Set{N}\big(\me_{k,1}^{(m)}\rmv, \me_{k,2}^{(m)}; [p_{k,1} \ist p_{k,2}]^{\mathrm{T}}\rmv\rmv\rmv,\sigma^{2}_{\text{p}}\V{I}_2\big)$.
Here, the variance $\sigma^{2}_{\text{p}}$ is obtained via moment matching by setting $\sigma^{2}_{\text{p}}$ to the variance of a uniform distribution defined by the side length of a data cell given by $1$m, i.e., $\sigma^{2}_{\text{p}} = 1/12$. 
In PMB and PMB-AM, $f_{1,\text{p}}(\me_{k,1}^{(m)}\rmv, \me_{k,2}^{(m)}|\st_k)$ is evaluated by setting $[\me_{k,1}^{(m)}\ist \me_{k,2}^{(m)}]^{\T}$ equal to the 2D cell center point. Furthermore, we model $f_{0,\text{p}}(\me_{k,1}^{(m)}\rmv, \me_{k,2}^{(m)})$ as uniformly distributed over the ROI.  It has been observed, that modeling the likelihood function $f_{1,\text{p}}(\me_{k,1}^{(m)}\rmv, \me_{k,2}^{(m)}|\st_k)$ constant  on the area of the corresponding data cell and zero otherwise leads to almost identical performance compared to the aforementioned Gaussian model.

\begin{figure}[t!]
 \vspace*{-0.5mm}
\centering
\footnotesize
{\scalebox{1}{\hspace{-2mm}\includegraphics[scale=0.7]{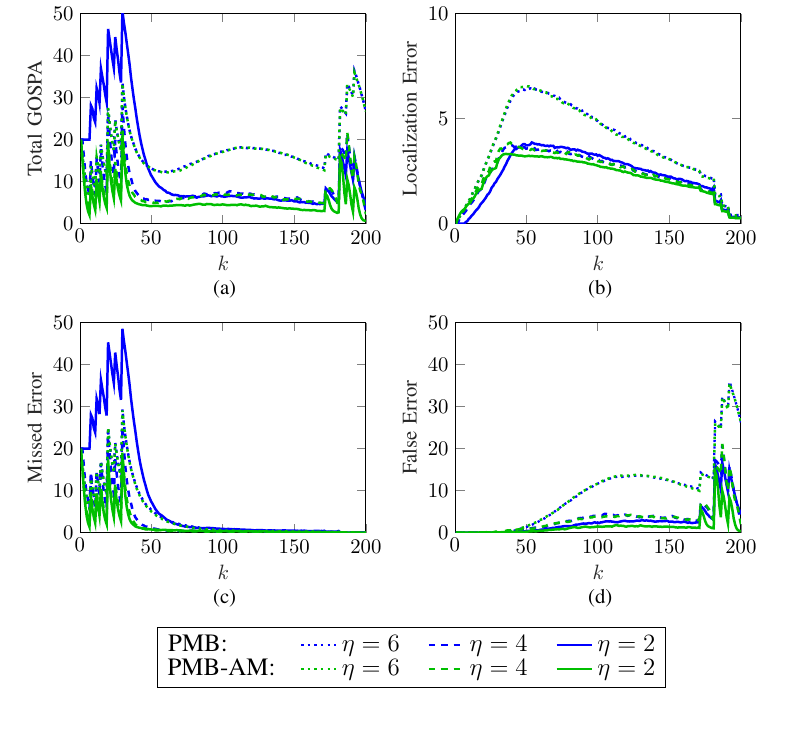}}}
\vspace{-10mm}
\caption{GOSPA error of PMB (blue curves) and PMB-AM (green curves) versus time $k$ for different detection thresholds $\eta$.}
\label{fig:GOSPA1} 
\vspace{-6mm}
\end{figure}

\subsection{Filter Settings and Simulation Results}

We employ particle implementations of PMB-CM, PMB-AM, and PMB. We represent the spatial pdf of each Bernoulli component by $3,\!000$ particles, the posterior PHD by $50,\!000$ particles, and the birth PHD by another $50,\!000$; the resulting $100,\!000$ particles are reduced to $50,\!000$ again after the update step.
More precisely, the birth PHD is given by $\PHD_{\text{B}}(\st_k) \rmv=\rmv \mu_{\text{B}}\ist f_{\text{B}}(\st_k)$ with $\mu_{\text{B}} \rmv=\rmv 5/32^2$ and $f_{\text{B}}(\st_k) \rmv= f(p_{k,1}, p_{k,2})\ist f_{\text{v}}\big(\dot{p}_{k,1},\dot{p}_{k,2}\big)\ist f_{\text{I}}(\gamma_{k})$ where $f(p_{k,1},p_{k,2})$ is uniform over the ROI, $f_{\text{v}}\big(\dot{p}_{k,1},\dot{p}_{k,2}\big)$ was defined in the previous subsection, and $f_{\text{I}}(\gamma_{k})$ is uniform from $0$ to $30$.
We set the survival probability to $\su \rmv=\rmv 0.999$ and the recycling threshold (cf. Section \ref{sec:AppUpd}) to $\eta_{\text{R}} \rmv=\rmv 0.1$.

The tracking performance is evaluated by means of the generalized optimal subpattern assignment (GOSPA) metric~ \cite{Rah17} with parameters $p \rmv=\rmv 1$, $c \rmv=\rmv 20$, and $\beta \rmv=\rmv 2$ for different thresholds $\eta \rrmv\in\rrmv \{2,4,6\}$. 
We show the GOSPA errors for PMB and PMB-AM in Fig.\!\! \ref{fig:GOSPA1}. It can be seen that the use of AM improves tracking performance.
In particular, as Fig.\! \ref{fig:GOSPA1}(c) indicates, surprisingly, the detection performance of newborn objects of PMB decreases for small $\eta$, which in turn also increases the total GOSPA error. This can be attributed to the fact that for low $\eta$ not only the possibility of object detection but also the number of clutter measurements increases. Here, AM helps the tracker to distinguish between object-generated measurements and clutter measurements.

\begin{figure}[t!]
 \vspace*{-0.5mm}
\centering
\footnotesize
{\scalebox{1}{\hspace{-2mm}\includegraphics[scale=0.7]{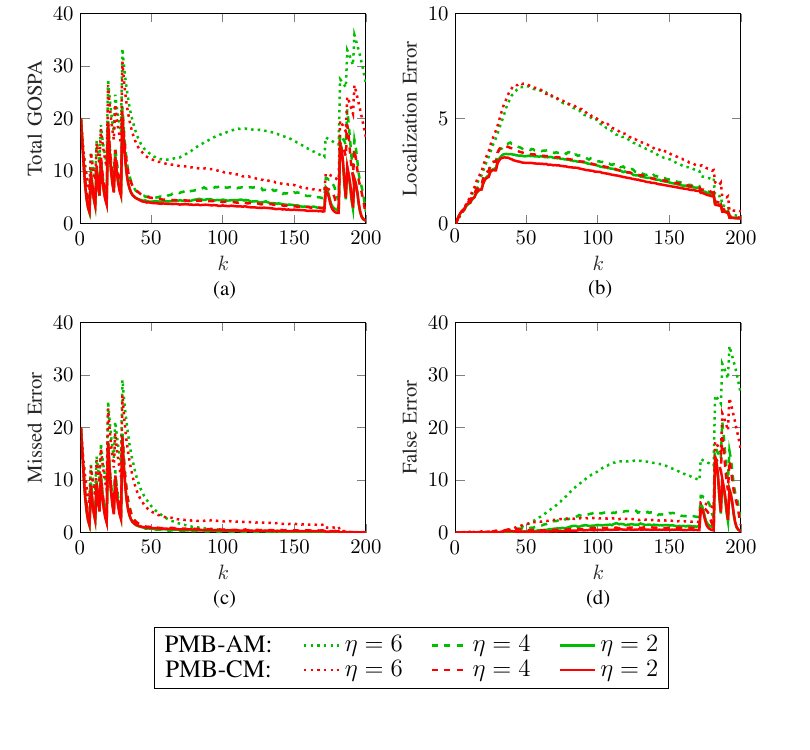}}}
\vspace{-10mm}
\caption{GOSPA error of PMB-AM (green curves) and proposed PMB-CM (red curves) versus time $k$ for different detection thresholds $\eta$.}
\label{fig:GOSPA2} 
\vspace{-2mm}
\end{figure}

Next, in Fig. \ref{fig:GOSPA2} we compare the GOSPA errors of PMB-AM and PMB-CM.
As can be seen there, PMB-CM outperforms PMB-AM for all three values of $\eta$.
This can be attributed to the applied point measurement approximation described in Section \ref{sec:PMM_wAI}, which seems to lead to a higher likelihood that a measurement was generated by a newborn object, which in turn to faster generation of new Bernoulli components.
As a result, PMB-AM detects newly appeared objects slightly faster, as shown in Fig. \ref{fig:GOSPA2}(c), but on the other hand produces a significant higher number of false tracks than PMB-CM, as shown in Fig. \ref{fig:GOSPA2}(d); this effect is even pronounced in scenarios with larger $\eta$.


Finally, we compare the three PMB filters also in terms of runtime/complexity; the average number of detections per time step $k$, i.e., $\overline{N}_{\text{Cell}} = |\Set{M}_k^{\text{D}}|$, and the average filter runtimes of PMB, PMB-AM, and PMB-CM for different values of $\eta$, obtained with a Matlab implementation on an Intel Xeon Gold 5222 CPU and measured using 1000 Monte Carlo runs, are reported in Table \ref{fig:RT}.
The table shows that PMB-AM is slightly slower than PMB due to the additional processing caused by AM. 
While the runtime of PMB-CM increases only slightly for lower values of $\eta$, the runtime increase of PMB and PMB-AM is significant.
We can thus conclude that the proposed PMB filter using thresholded cell measurements achieves a very good accuracy/complexity tradeoff.

\section{Conclusion}
We proposed a Poisson multi-Bernoulli (PMB) filter that uses thresholded cell measurements.
In a simulation experiment, we demonstrated the advantages of our proposed PMB filter using thresholded cell measurements compared to the conventional PMB filter with and without amplitude information. 
Future work includes advanced measurement models that account for multiple objects being in the same cell and objects contributing to multiple cells.
Further promising future research venues are the combination of probabilistic focalization with neural networks \cite{Zhang2024} and particle flow techniques~ \cite{Liang2024}.

\begin{table}
\centering
\vspace*{2mm}
{\small
\hspace*{0mm}	\begin{tabular}{|lllll|}	
	\hline
	\rule{0mm}{2.8mm}\!\! Threshold $\eta$  &\hspace{3mm} 6	  &\hspace{3mm}  4 &\hspace{3mm} 2 &\hspace{3mm} 0 \\[.12mm] \hline	
	\rule{0mm}{3.2mm}\hspace{-1.2mm} $\overline{N}_{\text{Cell}}$ & \hspace{1.5mm}1.28 & \hspace{3mm}3.51 & 143.15 & \hspace{1.5mm}1024  \\[.12mm] \hline	
	 \rule{0mm}{3.2mm}PMB & \hspace{1.5mm}6.95\ist s & \hspace{1.5mm}10.05\ist s & 183.65\ist s & \\[.1mm]	
	 PMB-AM  & \hspace{1.5mm}7.02{\ist}s & \hspace{1.5mm}10.32\ist s & 198.98\ist s & \\[.1mm]	
	 PMB-CM  & \hspace{1.5mm}9.52\ist s & \hspace{3mm}9.6{\ist}s &   \hspace{1.5mm}10.40{\ist}s &  \hspace{1.5mm}11.13{\ist}s \\[.1mm]
	\hline	
	\end{tabular}}

\vspace{1.5mm}
\caption{Average number of detections $\overline{N}_{\text{Cell}}$ per time step $k$ and average filter runtimes of PMB, PMB-AM, and PMB-CM for different values of $\eta$.}
\label{fig:RT} 
\vspace{-5mm}
\end{table}

\appendices
\renewcommand*\thesubsectiondis{\thesection.\arabic{subsection}}
\renewcommand*\thesubsection{\thesection.\arabic{subsection}}
\vspace{0mm}
\section{} 
\label{sec:App_A}
\vspace{-1mm}

\begin{figure*}
\vspace{-4.5mm}
\setcounter{equation}{39}
\begin{align} 
f(\RFSst_k|\mev_{1:k})\ist 
&\propto \sum_{\RFSst_{k}^{(1)} \uplus\ist \RFSst_{k}^{(2)} =\ist\RFSst_{k}} \sum_{\assv_k \in \Set{A}_k} \bigg(\prod_{j' \in\ist \Set{J}_{k-1}^{(0)}} 1 -\ex_{k|k-1}^{(j')} \bigg) \bigg(\prod_{j\ist\in\ist\Set{J}_{k-1}^{(1)}} \ex_{k|k-1}^{(j)}\ist \delta^{(\ass_k^{(i)})}(\st_k^{(i)})\ist f'\big(\me_k^{(\ass_k^{(i)})}|\st_k^{(i)}\big)\ist \sd_{k|k-1}^{(j)}(\st_k^{(i)}) \bigg)  \nn\\[1mm]
&\times \bigg(\prod_{\st_k^{(i')}\in\ist\RFSst_k^{(1)}} \delta^{(\ass_k^{(i')})}(\st_k^{(i')})\ist f'\big(\me_k^{(\ass_k^{(i')})}|\st_k^{(i')}\big)\ist \lambda_{k|k-1}(\st_k^{(i')}) \bigg)\ist \prod_{m \ist\in\ist \bar{\mathcal{M}}'_{\V{a}_k}} p_{\text{FA}}^{(m)}\ist f_{0,\eta}\big(\me_k^{(m)}\big)  \label{eq:UpdEx2}
\end{align}
\hrule\hrule
\vspace{-3mm}
\end{figure*}

In this appendix, we derive the exact update step applied in Section \ref{sec:ExUpd}.
The update step converts the predicted posterior pdf $f(\RFSst_k|\mev_{1:k-1})$ into the updated posterior pdf $f(\RFSst_{k}|\mev_{1:k})$ by applying Bayes theorem according to 
\vspace{0mm} 
\setcounter{equation}{34}
\begin{equation}
\label{eq:UpdEx1}
f(\RFSst_k|\mev_{1:k}) \propto f(\mev_k|\RFSst_k)\ist f(\RFSst_k|\mev_{1:k-1})\ist.
\vspace{0.5mm} 
\end{equation}
Recap that the predicted posterior pdf $f(\RFSst_k|\mev_{1:k-1})$ is of PMB form and given by expression \eqref{eq:PredPost}; 
it can be rewritten as
\begin{align}
\hspace{-1mm}f(\RFSst_k|\mev_{1:k-1})\ist &=\rrmv \sum_{\RFSst_{k}^{(1)} \uplus\ist \RFSst_{k}^{(2)} =\ist\RFSst_{k}}\hspace{-3mm} f^{\text{P}}\big(\RFSst_{k}^{(1)}\big) \prod_{j\in\ist\Set{J}_{k-1}} f^{(j)}\big(\RFSst_{k}^{(2,j)}\big) \nn \\[-2mm]
\label{eq:UpdEx_PredPost} \\[-5mm]
\nn
\end{align}
with $\RFSst_{k}^{(2)} \rmv=\rmv \RFSst_{k}^{(2,1)} \uplus\ldots\uplus \RFSst_{k}^{(2,J_{k-1})}$\rmv. 
Next, we rewrite $f(\mev_k,\assv_k|\RFSst_k)$ in \eqref{eq:Like2} according to
\vspace{0.5mm}
\begin{align}
f(\mev_k,\assv_k|\RFSst_k) &\propto \bigg( \prod_{i \ist\in\ist \Set{I}_{k}} \delta^{(\ass_k^{(i)})}\big(\st_k^{(i)}\big)\ist f'\big(\me_k^{(\ass_k^{(i)})}|\st_k^{(i)}\big) \bigg) \nn \\
&\hspace{3mm}\times \prod_{m \ist\in\ist \bar{\mathcal{M}}'_{\V{a}_k}}\rrmv p_{\text{FA}}^{(m)} f_{0,\eta}\big(\me_k^{(m)}\big) \label{eq:App_Like2} \\[-8mm]
\nn
\end{align}
with 
\vspace{0mm}
\begin{align}
&f'\big(\me_k^{(\ass_k^{(i)})}|\st_k^{(i)}\big) \nn \\[2mm]
&\hspace{2mm}=
\begin{cases}
\big(1 - p_{\text{D}}^{(\ass_k^{(i)})}\big(\st_k^{(i)}\big)\big)/\big(1 - p_{\text{FA}}^{(\ass_k^{(i)})}\big), &\rrmv \me_k^{(\ass_k^{(i)})} = \eta \\[2mm]
p_{\text{D}}^{(\ass_k^{(i)})}\big(\st_k^{(i)}\big)\ist f_{1,\eta}\big(\me_k^{(\ass_k^{(i)})}\big|\st_k^{(i)}\big), &\rrmv \me_k^{(\ass_k^{(i)})} > \eta\ist.
\vspace{0.5mm} 
\end{cases}
\label{eq:App_CondPDF_PI} \\[-5mm]
\nn
\end{align}
Here, we have used the decomposition of $\prod_{m \ist\in\ist \bar{\mathcal{M}}_{\V{a}_k}} f(\me_k^{(m)})$ into $\Big(\prod_{m \ist\in\ist \bar{\mathcal{M}}'_{\V{a}_k}} p_{\text{FA}}^{(m)}\ist f_{0,\eta}\big(\me_k^{(m)}\big)\Big)\ist\prod_{m' \ist\in\ist \bar{\mathcal{M}}''_{\V{a}_k}}  1 \rmv-\rmv p_{\text{FA}}^{(m')}$ and then formally multiplied and divided by $\prod_{m \ist\in\ist \mathcal{M}^{\text{D}}_k} 1 \rmv-\rmv p_{\text{FA}}^{(m)}$\rmv. Here, $\bar{\mathcal{M}}'_{\V{a}_k}$ and $\bar{\mathcal{M}}''_{\V{a}_k}$ comprise all cell indexes $m\rmv\in\rmv\bar{\Set{M}}_{\assv_k}$ with $\me_k^{(m)} \rmv= \eta$ and $\me_k^{(m)} \rmv> \eta$\ist, respectively.

Now, inserting \eqref{eq:UpdEx_PredPost} and \eqref{eq:App_Like2} together with \eqref{eq:Like1} into \eqref{eq:UpdEx1} yields the posterior pdf $f(\RFSst_k|\mev_{1:k})$ in \eqref{eq:UpdEx2}.
Here, $\Set{J}_{k-1}^{(0)}$ comprises all Bernoulli components $j \rmv\in\rmv \Set{J}_{k-1}$ with $\RFSst_k^{(2,j)} \rmv=\rmv \emptyset$ and $\Set{J}_{k-1}^{(1)}$ all Bernoulli components $j \rmv\in\rmv \Set{J}_{k-1}$ with $\RFSst_k^{(2,j)} \rmv= \{\st_k^{(i)}\}$.  
Rewriting \eqref{eq:UpdEx2} in terms of weighted Bernoulli pdfs as in \cite{Wil:J15,Kro21TBD} and using $\assvR'_k$ as defined in Section \ref{sec:ExUpd}, we finally obtain \eqref{eq:upd1}.

\renewcommand{\baselinestretch}{.95}
\selectfont
\bibliographystyle{IEEEtran}
\bibliography{references_Kr}

\end{document}